\documentclass[journal]{IEEEtran}

\ifCLASSINFOpdf
\else
\fi 

\usepackage{lipsum}
\usepackage[cmex10]{amsmath}
\usepackage{amssymb}   
\usepackage{amsxtra}
\usepackage{amscd}
\usepackage{amsthm}
 \usepackage{amsxtra}
 \usepackage{amscd}
 \usepackage{amsthm}
  \usepackage{bm}
\usepackage{tikz}
\usetikzlibrary{matrix}
\usepackage{stfloats}
\usepackage{graphicx}   
\usepackage{algorithm}
\usepackage{algpseudocode}
\usepackage{array}  
\usepackage{authblk}
\usepackage{multirow}  
\usepackage{arydshln}
\usepackage{stmaryrd}
\usepackage{booktabs}
\usepackage{placeins}
\usepackage{hyperref}
\hypersetup{%
  citebordercolor=blue
}
\usepackage{cite}

\newcommand{\lr}[1]{\left(#1\right)} 

\AtBeginDocument{%
    \setlength{\abovedisplayskip}{4pt plus 1pt minus 1pt}
    \setlength{\belowdisplayskip}{4pt plus 1pt minus 1pt}
    \setlength{\abovedisplayshortskip}{2pt plus 1pt minus 1pt}
    \setlength{\belowdisplayshortskip}{3pt plus 1pt minus 1pt}
    \setlength{\jot}{2pt}
}

\begin{document}
\bstctlcite{IEEEexample:BSTcontrol}
\title{\huge Ternary Noise Modulation\vspace{-0.2cm}}

\author{
  Ata Bilgin,~\IEEEmembership{Graduate Student Member,~IEEE,} Erkin Yapıcı,~\IEEEmembership{Graduate Student Member,~IEEE,} \vspace{-0.4cm}\\Yusuf Islam Tek,~\IEEEmembership{Graduate Student Member,~IEEE,}
  and Ertugrul Basar,~\IEEEmembership{Fellow,~IEEE}
\thanks{A. Bilgin and Y. I. Tek are with the Department of Electrical and Electronics Engineering, Koc University, Sariyer, Istanbul, Türkiye. Emails: abilgin20@ku.edu.tr and ytek21@ku.edu.tr. Y. I. Tek is also with the R\&D Department, Turk Telekom, 06030, Ankara, Türkiye. \\ Email: yusufislam.tek@turktelekom.com.tr}%
\thanks{E. Yapici is with the Department of Electrical and Electronics Engineering, Bogazici University, Istanbul, Türkiye. \\Email: erkin.yapici@std.bogazici.edu.tr}%
\thanks{E. Basar is with the Tampere Wireless Research Centre, Department of Electrical Engineering, Tampere University, 33720 Tampere, Finland, also holding an adjunct position at the Department of Electrical and Electronics Engineering, Koc University, Sariyer, Istanbul, Türkiye. Email: ertugrul.basar@tuni.fi}%

\thanks{This work is supported by the Scientific and Technological Research Council of Türkiye (TÜBİTAK) through the 1515 Frontier R\&D Laboratories Support Program for the Türk Telekom 6G R\&D Lab (Project No. 5249902) and under Grant No. 124E146.}
\vspace{-1.2cm}}


\maketitle

\begin{abstract}
By exploiting noise as an information-bearing resource, noise-driven communication offers a promising framework for low-complexity wireless system design. In this letter, the scheme of ternary noise modulation (T-NoiseMod) is proposed for noise-based wireless communication scenarios, where information is encoded into the statistical characteristics of artificial noise. Unlike conventional binary NoiseMod, which employs two variance levels, the proposed scheme introduces a third transmission state: intentional silence. By pairing two consecutive noise blocks, the signaling scheme is expanded to eight valid state combinations, enabling the transmission of three information bits per signaling interval. In our proposed scheme, the two-stage receiver is developed, consisting of mean-based silent-state detection followed by variance-based low/high classification. An approximate analytical expression for the bit error probability (BEP) is derived for Rayleigh fading. Our computer simulation results match closely with our approximate theoretical results and show the effects of key system parameters. Furthermore, comparisons with binary NoiseMod, Q-NoiseMod, and OODN demonstrate the inherent trade-off between reliability and rate.
\end{abstract}

\vspace{-0.4cm}

\IEEEpeerreviewmaketitle

\section{Introduction}
Next-generation wireless networks are expected to support massive machine-type communications and ultra-low-power connectivity under highly dynamic environments. These ambitious requirements call for fundamentally new transmission paradigms that move beyond waveform-centric designs and strict synchronization assumptions \cite{butt2023ambientiotmissinglink}. Noise-based communication has emerged as an attractive alternative to conventional modulation techniques for wireless systems that prioritize low complexity. Instead of encoding information onto deterministic waveforms, noise-based schemes embed data into the statistical properties of artificially generated noise signals \cite{basar2023thermal, 10373568}. This statistical signaling paradigm allows for non-coherent receiver designs, particularly in fading environments. Moreover, noise-like transmissions are promising for energy-constrained Internet-of-Things applications due to their statistical signaling structure and low-complexity receiver potential \cite{da2026survey,dos2025off, yapici2025noisemodulationwirelessenergy, anand2019wireless, StochasticShiftKeying}. Recent IoT-oriented wireless access studies have also emphasized the importance of efficient transmission design for future wireless applications \cite{Lin2021RSMAIoT}.

In early foundational studies, thermal noise modulation (TherMod) has been introduced as a means of carrying information through inherent or artificially generated noise processes, establishing the feasibility of noise-driven communication without deterministic waveforms \cite{da2026survey,kish2005stealth}. Moreover, existing noise modulation techniques predominantly rely on binary signaling structures, where information is conveyed through two distinct statistical states, such as low and high noise variance or different mean levels \cite{ yapici2025noisemodulationwirelessenergy,dos2025off,11242237,ma2025initial}. Building upon this concept, subsequent binary noise modulation schemes demonstrated that reliable communication can be achieved by exploiting simple statistical properties of noise, enabling low-complexity receiver architectures that rely on statistical detection without requiring explicit phase or frequency synchronization. While these binary schemes offer simple implementation and robust detection performance, they fundamentally limit the achievable information rate per signaling interval, leading to inherently limited spectral efficiency. To extend noise-based communication to multi-user scenarios, noise-domain non-orthogonal multiple access (ND-NOMA) has been proposed, where multiple users embed their information into independent statistical dimensions of noise, such as the mean, variance, or correlation \cite{yapici2024noisedomainnonorthogonalmultipleaccess}. In addition, on–off digital noise modulation (OODN) was introduced as a binary noise-based scheme that conveys information through the presence or absence of digitally generated noise to achieve better performance \cite{dos2025off}. However, in OODN, silence is used within a binary presence/absence signaling mechanism. In contrast, T-NoiseMod incorporates silence as a third signaling state and jointly exploits it with low- and high-variance noise levels. This enlarges the signaling alphabet while preserving the variance-based signaling principle of NoiseMod, resulting in a different rate--reliability tradeoff.

Despite these advances, existing binary noise modulation and ND-NOMA frameworks except OODN scheme, still rely on two-level statistical signaling and do not explicitly exploit silence as an information-bearing state. In most designs, silence is treated as an idle or unused resource rather than as a deliberate signaling dimension, thereby limiting the achievable trade-off between information rate and reliability. This observation motivates the exploration of higher-order noise-based signaling structures that can expand the signaling alphabet without sacrificing the intrinsic advantages of noise-domain communication. 

To this end, this letter introduces ternary noise modulation (T-NoiseMod), a noise-based signaling framework that incorporates an intentional silent state alongside low- and high-variance noise transmissions. By pairing two consecutive noise blocks, the proposed ternary structure enables the transmission of three information bits per extended signaling interval. Furthermore, a two-stage receiver architecture is developed, consisting of mean-based silent-state detection followed by variance-based low/high discrimination, enabling higher information efficiency and a flexible rate–reliability tradeoff.

\vspace{-0.4cm}
\section{System Model}

\begin{figure*}[t]
    \centering
    \includegraphics[width=0.9\textwidth]{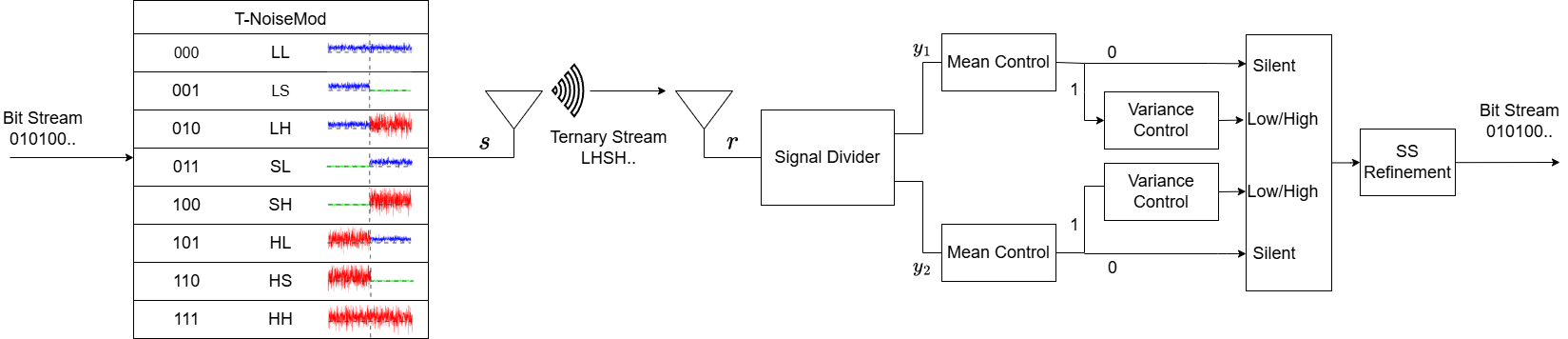}
    \vspace{-0.2cm}
    \caption{The transceiver model of the proposed T-NoiseMod scheme.}
    \vspace{-0.4cm}
    \label{fig:system_model}
\end{figure*}

In this section, we present the proposed T-NoiseMod scheme, which is illustrated in Fig. \ref{fig:system_model}. Consider a noise-based wireless communication system in which users embed information into the statistical characteristics of an artificial noise signal. In contrast to conventional binary NoiseMod schemes, which use two distinct noise variances or mean levels (low/high) \cite{10373568,yapici2025noisemodulationwirelessenergy}, we propose a ternary noise-based signaling mechanism, T-NoiseMod, consisting of two consecutive blocks. Each transmission block can be in one of three possible states: low variance ($\textit{L}$), high variance ($\textit{H}$), or silent ($\textit{S}$). The silent state \textit{S} is an intentional signaling option that conveys information through non-transmission, similar to backscatterer systems. In addition to enlarging the signaling alphabet without additional waveform resources, the silent state can reduce the average transmit activity by allowing information to be conveyed through non-transmission in one of the two blocks. By pairing two consecutive blocks, a total of eight distinct symbol combinations are formed, allowing the transmission of three information bits per signaling pair.

Each bit stream is grouped into triplets \([b_1,b_2,b_3] \in \{0,1\}^3\), which are mapped onto two consecutive noise blocks with $N$ elements according to a predefined distributions mapping table. The corresponding pair of transmission states is denoted by 
\((s_1, s_2) \in \{L, S, H\}\), 
where each element represents the state of the first and second block, respectively. The noise blocks can be denoted as $\mathbf{x}^i=[x^i[1],\dots,x^i[k],\dots,x^i[N]]$, where $i \in \{1,2\}$ corresponds to the block number (i.e., first or second block). Each block may consist of one of the following Gaussian processes, depending on its state:
\begin{equation}
{x}^i[k] \sim
\begin{cases}
\mathcal{CN}(\mu, \sigma_L^2), & \text{if state is low (\textit{L}),}\\[2pt]
\mathcal{CN}(\mu, \sigma_H^2), & \text{if state is high (\textit{H}),} \\[2pt]
0, & \text{if state is silent (\textit{S})},
\end{cases}
\end{equation}
where $\sigma^2_L$, $\sigma^2_H$, and $\mu$ denote the low and high variances of the proper complex Gaussian noise samples, and the real-valued reference mean value used for silent-state discrimination, respectively. Note that $\mathbf{0}_N$ represents a vector with all zero elements, and \textit{S} does not represent an \textit{idle} period outside the signaling structure; instead, it is an intentional transmission state used for bit encoding. Hence, even though no samples are transmitted in a silent block, the choice of \textit{S} carries information and affects the receiver's decision process. An exemplary mapping table is given in Table \ref{tab:ternary_mapping}. 

After generating blocks according to the corresponding bits, the complete transmit block $\mathbf{s}=\left[ \mathbf{x}_1, \, \mathbf{x}_2 \right]\in\mathbb{C}^{1\times2N}$ is sent over the wireless channel and received as $\mathbf{r}\in\mathbb{C}^{1\times2N}$ as $\mathbf{r} = h\mathbf{s} + \mathbf{w}$, where $h\sim\mathcal{CN}(0,1)$ is the complex baseband Rayleigh channel coefficient and $\mathbf{w}$ is the vector of additive white Gaussian noise (AWGN) samples, whose elements follow $\mathcal{CN}(0,\sigma^2_w)$. For the subsequent performance analysis, we introduce the parameter \( \delta = \sigma_L^2 / \sigma_w^2 \), which quantifies the variance ratio between the useful signal and the AWGN. This parameter serves as an analogue to the signal-to-noise ratio (SNR) commonly used in conventional communication systems. An increase in \( \delta \) implies improved receiver robustness against additive noise perturbations, thereby resulting in a reduced BEP.

At the receiver, the incoming $2N$-sample vector $\mathbf{r}$ is divided into two consecutive segments, 
denoted by $\mathbf{y}_1$ and $\mathbf{y}_2$, each corresponding to one transmitted block of length $N$ and
that is,
$
\mathbf{y}_i = [y_i[1], y_i[2], \dots, y_i[N]], \quad i \in \{1,2\}.
$
Each segment $\mathbf{y}_i$ is then processed independently to determine its state $\hat{s}_i \in \{L, S, H\}$ through a two-stage detection process described below. Note that the instantaneous channel gain \(h\) is assumed to remain quasi-static over the complete T-NoiseMod signaling pair, i.e., over the two consecutive length-\(N\) sub-blocks, and is perfectly known at the receiver.

\textbf{1) Mean-based silent detection:}
The receiver computes the sample mean of each segment as
\begin{equation}
    \bar{y}_i = \frac{1}{N}\sum_{k=1}^{N} y_i[k].
    \label{eq:mean}
\end{equation}
    The value $\bar{y}_i$ is compared with the reference mean $h\mu$ to determine whether the block is silent or not. 
    If $|\bar{y}_i| < |\,\bar{y}_i - h\mu\,|$, the segment is declared as silent, i.e., $\hat{s}_i = S$.
    Otherwise, the segment is considered active and processed in the next step.

\begin{table}[t]
\renewcommand{\arraystretch}{1.5}
\centering
\caption{An exemplary ternary mapping table.}
\vspace{-0.2cm}
\label{tab:ternary_mapping}
\begin{tabular}{|c|c|c|}
\hline
\begin{tabular}[c]{@{}c@{}}$[b_1, \, b_2, \, b_3]$\end{tabular} & \begin{tabular}[c]{@{}c@{}}$(s_1,s_2)$\end{tabular} & $\mathbf{s}=[\mathbf{x}_1, \, \mathbf{x}_2]$ \\ \hline
$[0 \, 0 \, 0]$ & \textit{(L,L)} & $\left[\sim\mathcal{CN}(\mu, \sigma_L^2\mathbf{I}_N), \, \sim\mathcal{CN}(\mu, \sigma_L^2\mathbf{I}_N)\right]$ \\ \hline
$[0 \, 0 \, 1]$ & \textit{(L,S)} & $\left[\sim\mathcal{CN}(\mu, \sigma_L^2\mathbf{I}_N), \, \mathbf{0}_N\right]$ \\ \hline
$[0 \, 1 \, 0]$ & \textit{(L,H)} & $\left[\sim\mathcal{CN}(\mu, \sigma_L^2\mathbf{I}_N), \, \sim\mathcal{CN}(\mu, \sigma_H^2\mathbf{I}_N)\right]$ \\ \hline
$[0 \, 1 \, 1]$ & \textit{(S,L)} & $\left[\mathbf{0}_N, \, \sim\mathcal{CN}(\mu, \sigma_L^2\mathbf{I}_N)\right]$ \\ \hline
$[1 \, 0 \, 0]$ & \textit{(S,H)} & $\left[\mathbf{0}_N, \, \sim\mathcal{CN}(\mu, \sigma_H^2\mathbf{I}_N)\right]$ \\ \hline
$[1 \, 0 \, 1]$ & \textit{(H,L)} & $\left[\sim\mathcal{CN}(\mu, \sigma_H^2\mathbf{I}_N), \, \sim\mathcal{CN}(\mu, \sigma_L^2\mathbf{I}_N)\right]$ \\ \hline
$[1 \, 1 \, 0]$ & \textit{(H,S)} & $\left[\sim\mathcal{CN}(\mu, \sigma_H^2\mathbf{I}_N), \, \mathbf{0}_N\right]$ \\ \hline
$[1 \, 1 \, 1]$ & \textit{(H,H)} & $\left[\sim\mathcal{CN}(\mu, \sigma_H^2\mathbf{I}_N), \, \sim\mathcal{CN}(\mu, \sigma_H^2\mathbf{I}_N)\right]$ \\ \hline
\end{tabular}
\vspace{-0.5cm}
\end{table}

\textbf{2) Variance-based state detection:}
    For non-silent segments, the empirical variance is estimated as
    \begin{equation}
        \widehat{\sigma}_{y_i}^2 = \frac{1}{N-1}\sum_{k=1}^{N} |y_i[k]-\bar{y}_i|^2.
        \label{eq:variance}
    \end{equation}
    The estimated variance $\widehat{\sigma}_{y_i}^2$ is then compared against a threshold $\tau$, 
    derived from a log-likelihood ratio test \cite{OPTIMUM_VAR_ML_THRESHOLD_2024}:
    \begin{equation}
        \tau = 
        \ln\lr{\!\frac{\sigma^2_{y_1}}{\sigma^2_{y_0}}}
        \frac{\sigma^2_{y_1}\sigma^2_{y_0}}{\sigma^2_{y_1}-\sigma^2_{y_0}},
        \label{eq:threshold}
    \end{equation}
    where 
    $\sigma^2_{y_0} = |h|^2\sigma_L^2 + \sigma_w^2$ and 
    $\sigma^2_{y_1} = |h|^2\sigma_H^2 + \sigma_w^2$.
    If $\widehat{\sigma}_{y_i}^2 > \tau$, the segment is classified as high-variance ($\hat{s}_i = H$);
    otherwise, it is classified as low-variance ($\hat{s}_i = L$). This threshold is used only after a segment is declared active by the mean-based detector.

An important design choice of the proposed T-NoiseMod signaling is that the $(S,S)$ state is not included in the valid symbol set, since at least one of the two blocks is kept active to provide an observable signaling interval and to form an eight-symbol alphabet carrying three information bits. However, due to noise fluctuations, the silent-state detector may occasionally classify both blocks as silent. To avoid this illegal decision, an additional consistency step is introduced. Specifically, if both blocks are classified as silent, we compute the active-reference distance
\(d_i=|\bar{y}_i-h\mu|^2\), \(i\in\{1,2\}\). The block with the smaller \(d_i\), i.e., the block whose sample mean is closer to the active reference \(h\mu\), is reinterpreted as an active block and is subsequently processed by the variance detector. This simple rule ensures that the $(S,S)$ case is avoided and guarantees that at least one block is treated as active during detection. After the detection process, the pair of estimated states $(\hat{s}_1,\hat{s}_2)\in\{L,S,H\}$ is mapped back to the corresponding bit triplet $[b_1,b_2,b_3]$ according to the transmitter mapping rule provided in Table~\ref{tab:ternary_mapping}.

\vspace{-3mm}
\section{Theoretical Analysis} 
\vspace{-2mm}
In this section, we derive the conditional and average BEP of the proposed T-NoiseMod scheme over quasi-static Rayleigh fading. Since each segment $\mathbf{y}_i$ 
is first processed independently, we analyze the preliminary single-segment 
state-detection error probabilities conditioned on $h$. The pairwise BEP is 
then obtained by combining these preliminary probabilities with the SS-consistency 
refinement. For compactness, we define the variance ratio $\alpha \triangleq \sigma_H^2/\sigma_L^2>1$ and recall $\delta=\sigma_L^2/\sigma_w^2$.

\vspace{-2mm}
\subsection{Conditional statistics of the sample mean}
For a given segment index $i\in\{1,2\}$, the sample mean $\bar{y}_i$ is defined as in \eqref{eq:mean}. Using $\mathbf{y}_i = h\mathbf{x}_i + \mathbf{w}_i$ and the definitions in Section~II, the distribution of $\bar{y}_i$ conditioned on the transmitted state is proper complex Gaussian:
\begin{equation}
\begin{aligned}
\bar{y}_i \, \big| \, S &\sim \mathcal{CN}\!\left(0,\frac{\sigma_w^2}{N}\right),\\
\bar{y}_i \, \big| \, L &\sim \mathcal{CN}\!\left(h\mu,\frac{|h|^2\sigma_L^2+\sigma_w^2}{N}\right),\\
\bar{y}_i \, \big| \, H &\sim \mathcal{CN}\!\left(h\mu,\frac{|h|^2\sigma_H^2+\sigma_w^2}{N}\right).
\end{aligned}
\label{eq:ybar_dist}
\end{equation}
\vspace{-0.7cm}
\subsection{Mean-based silent detection error probabilities}
\vspace{-1mm}
The silent-state decision follows a low-complexity closest-reference-mean rule between the silent reference mean zero and the active reference mean $h\mu$, which is ML-inspired under the effective mean-detection model. The mean-based rule in Section~II decides $S$ if $|\bar{y}_i|<|\bar{y}_i-h\mu|$. This is equivalent to the real-valued test
\begin{equation}
\Re\!\left\{\bar{y}_i (h\mu)^{\!*}\right\} \;\underset{\hat{s}_i = S}{\overset{\hat{s}_i \neq S}{\gtrless}}\; \frac{|h\mu|^2}{2}.
\end{equation}
Let $Q(\cdot)$ denote the Gaussian $Q$-function. Conditioned on $h$, the false-alarm probability (declaring active when $S$ is transmitted) is given by
\begin{equation}
p_{S\to A}(h) \triangleq \Pr(\hat{s}_i\neq S \mid S,h) =Q\!\left(\frac{|h\mu|\sqrt{N}}{\sqrt{2\sigma_w^2}}\right).
\label{eq:pSA}
\end{equation}
Similarly, the miss-detection probabilities (declaring $S$ when $L$ or $H$ is transmitted) are
\begin{align}
p_{L\to S}(h) &\triangleq \Pr(\hat{s}_i= S \mid L,h) =Q\!\left(\frac{|h\mu|\sqrt{N}}{\sqrt{2\sigma_{y_0}^2}}\right), \label{eq:pLS}\\
p_{H\to S}(h) &\triangleq \Pr(\hat{s}_i= S \mid H,h) =Q\!\left(\frac{|h\mu|\sqrt{N}}{\sqrt{2\sigma_{y_1}^2}}\right), \label{eq:pHS}
\end{align}

\vspace{-5.8mm}
\subsection{Variance-based low/high detection error probabilities}
\vspace{-1mm}
For segments declared active, the receiver computes $\widehat{\sigma}^2_{y_i}$ in \eqref{eq:variance}
and compares it to the likelihood ratio test (LRT) threshold $\tau$ in \eqref{eq:threshold}. Due to the $(S,S)$-consistency refinement, the final pair decision couples the two blocks, and obtaining exact closed-form expressions for the resulting state distribution is analytically intractable. In order to derive tractable expressions, we approximate the empirical variance by a Gaussian random variable (accurate for moderate/large $N$) as $\widehat{\sigma}^2_{y_i} \approx \mathcal{N}\!\left(\sigma_y^2,\frac{\sigma_y^4}{N-1}\right)$, and approximate the refinement by assuming that a truly active block is selected when the preliminary output is $(S,S)$, with equal selection probability for the two-active pairs. These approximations become increasingly accurate as $N$ grows and are validated by the close agreement between analytical and simulation results. Conditioned on $h$, the conditional confusion probabilities between $L$ and $H$ are then
\begin{align}
p_{L\to H}(h) &\triangleq \Pr(\widehat{\sigma}^2_{y_i}>\tau \mid L,h,\hat{s}_i\neq S) \nonumber\\
&=Q\!\left(\frac{\tau-\sigma_{y_0}^2}{\sigma_{y_0}^2/\sqrt{N-1}}\right), \label{eq:pLH}\\
p_{H\to L}(h) &\triangleq \Pr(\widehat{\sigma}^2_{y_i}\le\tau \mid H,h,\hat{s}_i\neq S) \nonumber\\
&=1-Q\!\left(\frac{\tau-\sigma_{y_1}^2}{\sigma_{y_1}^2/\sqrt{N-1}}\right). \label{eq:pHL}
\end{align}
Moreover, when a silent segment is mistakenly declared active, the same variance test produces $L$ or $H$. Since $y_i[k]|S \sim \mathcal{CN}(0,\sigma_w^2)$, we similarly obtain
\begin{equation}
p_{S\to H}(h) \triangleq \Pr(\widehat{\sigma}^2_{y_i}>\tau \mid S,h,\hat{s}_i\neq S) =Q\!\left(\frac{\tau-\sigma_w^2}{\sigma_w^2/\sqrt{N-1}}\right),
\label{eq:pSH}
\end{equation}
and $p_{S\to L}(h)=1-p_{S\to H}(h)$ under the event $\hat{s}_i\neq S$.

From the above expressions, we can calculate the single-segment ternary-state transition probabilities.
Combining the two stages yields the full transition probabilities $\Pr(\hat{s}_i=t \mid s_i=u,h)$ for $u,t\in\{S,L,H\}$. If $S$ is transmitted:
\begin{equation}
\begin{aligned}
\Pr(\hat{s}_i=S\mid S,h) &= 1-p_{S\to A}(h),\\
\Pr(\hat{s}_i=H\mid S,h) &= p_{S\to A}(h)\,p_{S\to H}(h),\\
\Pr(\hat{s}_i=L\mid S,h) &= p_{S\to A}(h)\,\big(1-p_{S\to H}(h)\big).
\end{aligned}
\end{equation}
If $L$ is transmitted:
\begin{equation}
\begin{aligned}
\Pr(\hat{s}_i=S\mid L,h) &= p_{L\to S}(h),\\
\Pr(\hat{s}_i=H\mid L,h) &= \big(1-p_{L\to S}(h)\big)\,p_{L\to H}(h),\\
\Pr(\hat{s}_i=L\mid L,h) &= \big(1-p_{L\to S}(h)\big)\,\big(1-p_{L\to H}(h)\big).
\end{aligned}
\end{equation}
If $H$ is transmitted:
\begin{equation}
\begin{aligned}
\Pr(\hat{s}_i=S\mid H,h) &= p_{H\to S}(h),\\
\Pr(\hat{s}_i=L\mid H,h) &= \big(1-p_{H\to S}(h)\big)\,p_{H\to L}(h),\\
\Pr(\hat{s}_i=H\mid H,h) &= \big(1-p_{H\to S}(h)\big)\,\big(1-p_{H\to L}(h)\big).
\end{aligned}
\end{equation}

For pairwise decision probability and conditional BEP, let
$(s_1,s_2)\in\mathcal{S}$ denote the transmitted state pair and let
$(\tilde{s}_1,\tilde{s}_2)$ denote the preliminary detected pair before
the consistency refinement. Since $\mathbf{y}_1$ and $\mathbf{y}_2$ depend
on disjoint sample sets and are first processed independently, the
preliminary pairwise probability can be written as
\begin{equation}
\Pr\!\big((\tilde{s}_1,\tilde{s}_2)\mid(s_1,s_2),h\big)
=
\Pr(\tilde{s}_1\mid s_1,h)\Pr(\tilde{s}_2\mid s_2,h).
\label{eq:prelim_pair_prob}
\end{equation}

The final detector output $(\hat{s}_1,\hat{s}_2)$ is obtained from
$(\tilde{s}_1,\tilde{s}_2)$ through the $(S,S)$ consistency refinement described
in Section~II. Therefore, the exact final pairwise probability is not, in general,
in a simple product form. Under the tractable refinement approximation stated above, let
$\Pi_{ab}(h\mid s_1,s_2)\triangleq
\Pr(\tilde{s}_1=a,\tilde{s}_2=b\mid s_1,s_2,h)$
denote the preliminary joint probability. Whenever the illegal event
$(\tilde{s}_1,\tilde{s}_2)=(S,S)$ occurs, its probability mass is
approximately redistributed over the valid outputs by assuming that a truly active block is selected.
For notational simplicity, let
$\Pi_{uv}\triangleq \Pi_{uv}(h\mid s_1,s_2)$,
$\Pi_{SS}\triangleq \Pi_{SS}(h\mid s_1,s_2)$, and
$\omega_{uv}\triangleq \omega_{uv}^{(s_1,s_2)}(h)$.
Accordingly, for $(u,v)\in\mathcal{S}$, the final transition probability
is approximated as
\begin{equation}
\begin{aligned}
\Pr\!\big((\hat{s}_1,\hat{s}_2)=(u,v)\mid(s_1,s_2),h\big)
\approx \Pi_{uv} + \Pi_{SS}\,\omega_{uv}.
\end{aligned}
\label{eq:final_pair_prob}
\end{equation}
where $\omega_{uv}$ denotes the approximate redistribution weight and satisfies
$\sum_{(u,v)\in\mathcal{S}}\omega_{uv}=1$.
For example, when $(s_1,s_2)=(L,S)$, the approximation assumes that the first block is selected as active and is finally classified as either $L$ or $H$, which gives
$\omega_{LS}^{(L,S)}(h)=1-p_{L\to H}(h)$ and
$\omega_{HS}^{(L,S)}(h)=p_{L\to H}(h)$. Similarly, for
$(s_1,s_2)=(S,H)$, we obtain
$\omega_{SH}^{(S,H)}(h)=1-p_{H\to L}(h)$ and
$\omega_{SL}^{(S,H)}(h)=p_{H\to L}(h)$. For the symmetric
two-active pairs $(L,L)$ and $(H,H)$, either active block is selected with probability $1/2$ under the approximation. The remaining cases are handled
in the same manner. The Monte Carlo simulations, however, implement the actual distance-based selection rule in Section~II.

Denote the set of valid state pairs by
$\mathcal{S}\triangleq\{(L,L),(L,S),(L,H),(S,L),(S,H),(H,L),(H,S),(H,H)\}$,
and let ${b}(s_1,s_2)\in\{0,1\}^3$ be the 3-bit label associated with
$(s_1,s_2)$ according to Table~\ref{tab:ternary_mapping}. The conditional
BEP given $h$ is then
\begin{equation}
\begin{aligned}
P_b(h)
&=
\frac{1}{|\mathcal{S}|}\sum_{(s_1,s_2)\in\mathcal{S}}
\sum_{\substack{(\hat{s}_1,\hat{s}_2)\in\mathcal{S}\\(\hat{s}_1,\hat{s}_2)\neq(s_1,s_2)}}
\frac{d_H\!\big({b}(s_1,s_2),{b}(\hat{s}_1,\hat{s}_2)\big)}{3} \\
&\hspace{2.2em}\times
\Pr\!\big((\hat{s}_1,\hat{s}_2)\mid(s_1,s_2),h\big),
\end{aligned}
\label{eq:Pb_cond}
\end{equation}
where $d_H(\cdot,\cdot)$ is the Hamming distance and the final transition
probabilities are approximated by \eqref{eq:final_pair_prob}.

Under quasi-static Rayleigh fading, $h\sim\mathcal{CN}(0,1)$ and $u\triangleq|h|^2$ is exponentially distributed with PDF $f_U(u)=e^{-u}$, $u\ge 0$. The unconditional average BEP is obtained by
\begin{equation}
\bar{P}_b = \mathbb{E}_{h}\!\left[P_b(h)\right] = \int_{0}^{\infty} P_b(\sqrt{u})\, e^{-u}\,du,
\label{eq:Pb_rayleigh_avg}
\end{equation}
where $P_b(\sqrt{u})$ denotes \eqref{eq:Pb_cond} evaluated at $|h|=\sqrt{u}$. We numerically evaluate the integral in \eqref{eq:Pb_rayleigh_avg} with the
adaptive quadrature method \cite{AdaptiveQuadrature}.

\section{Simulation Results} 
\begin{figure}[t]
    \centering
    \includegraphics[width=0.7\linewidth]{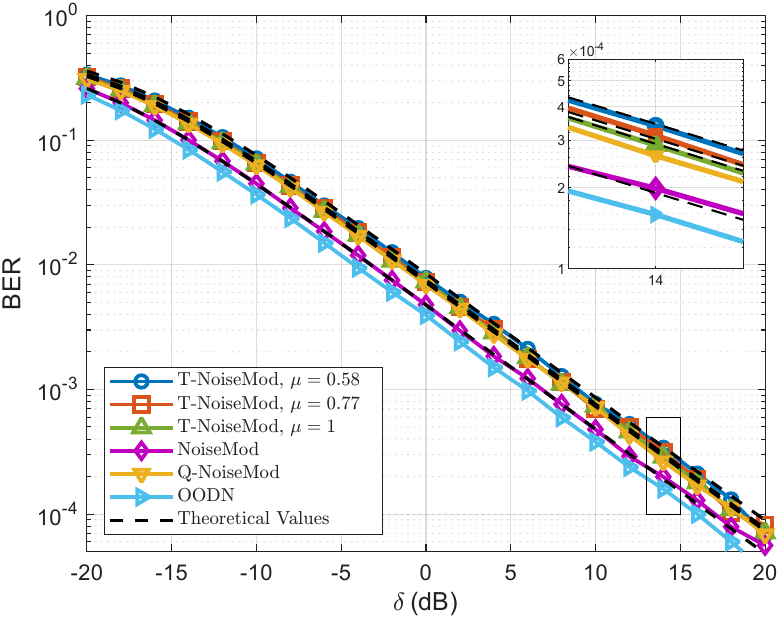}
    \vspace{-0.4cm}
    \caption{BER performance of conventional NoiseMod 
($N_B = 400 = 2N_T$), Q-NoiseMod ($N_Q = 400$), 
OODN ($N_O = 400$), and T-NoiseMod ($N_T = 200$) 
for varying $\mu$ values ($\alpha = 10$).}

\vspace{-0.4cm}
    \label{fig:mu_comp}
\end{figure}

In this section, we performed computer simulations to investigate the impact of the design parameters on the error performance of the proposed T-NoiseMod and validate the derived approximate analytical BEP expressions. Also, T-NoiseMod has been compared with the conventional NoiseMod benchmark. 
To clarify the normalization, $N$ in T-NoiseMod denotes the length of each sub-block. Therefore, a T-NoiseMod symbol occupies $2N$ samples and conveys three bits. Consequently, comparing T-NoiseMod with $N=200$ and binary NoiseMod with $N=400$ equalizes the total observation length per signaling decision. The average-power comparison depends on $\mu$. Throughout, the transmit power is normalized to unity, i.e., $\sigma_L^{2}=2/(1+\alpha)$ and $\sigma_H^{2}=2\alpha/(1+\alpha)$, so that $\sigma_L^{2}+\sigma_H^{2}=2$, while $\sigma_w^{2}=\sigma_L^{2}/\delta$ varies with $\delta$. Since each block is active with probability $3/4$ under equiprobable symbols, the average transmit power of T-NoiseMod is $\bar{P}_{\mathrm{T}}(\mu)=\frac{3}{4}\mu^{2}+\frac{3}{8}(\sigma_L^{2}+\sigma_H^{2})$, whereas NoiseMod and Q-NoiseMod operate at $\frac{1}{2}(\sigma_L^{2}+\sigma_H^{2})$. Hence, $\mu=1/\sqrt{3}\approx0.58$ equalizes the average transmit powers, so that the $\mu=0.58$ curve in Fig.~\ref{fig:mu_comp} is obtained under both an approximately common average power and a common observation length, while $\mu=0.77$ and $\mu=1$ correspond to $1.19$ and $1.5$ times the NoiseMod average power. For $\mu=1$, the same-$N$ comparisons correspond to equal-energy-per-bit operation, whereas the doubled-$N$ binary case provides an equal-observation-length comparison. Together, these results separate the impact of estimator averaging from the rate advantage of T-NoiseMod.

Fig.~\ref{fig:mu_comp} shows the BEP performance for different silent reference mean values and comparison with OODN and Q-NoiseMod. The approximate theoretical curves are obtained from \eqref{eq:Pb_rayleigh_avg} for $\mu\in\{0.58,0.77,1\}$. The benchmark Q-NoiseMod conveys two bits per complex symbol by independently assigning low/high variances to the in-phase and quadrature noise components \cite{11242237}, whereas OODN uses a silent state instead of low variance \cite{dos2025off}, different than classical NoiseMod. All schemes use the same number of samples within a transmission frame. OODN attains the lowest BEP among the considered schemes, as its binary on--off structure devotes the entire power contrast to the silence/activity decision. However, OODN conveys a single bit per $400$ samples, whereas Q-NoiseMod and T-NoiseMod convey two and three bits, respectively, over the same interval. Hence, T-NoiseMod trades a modest BEP degradation for a threefold rate advantage over OODN under identical average power and observation time. Furthermore, as observed from the results, increasing $\mu$ improves the mean-based silent/active discrimination by increasing the separation between the silent reference mean zero and the active reference mean $h\mu$, thereby reducing the overall BEP. In particular, small $\mu$ values are more vulnerable to silent/active confusion during deep fades, whereas moderate-to-large $\mu$ values provide improved reliability.
Also, the probability of the illegal $(S,S)$ detector output is low, and the proposed consistency refinement correctly resolves the majority of such events. Consequently, its impact on the overall BEP remains negligible.

\begin{figure}[t]
    \centering
    \includegraphics[width=0.7\linewidth]{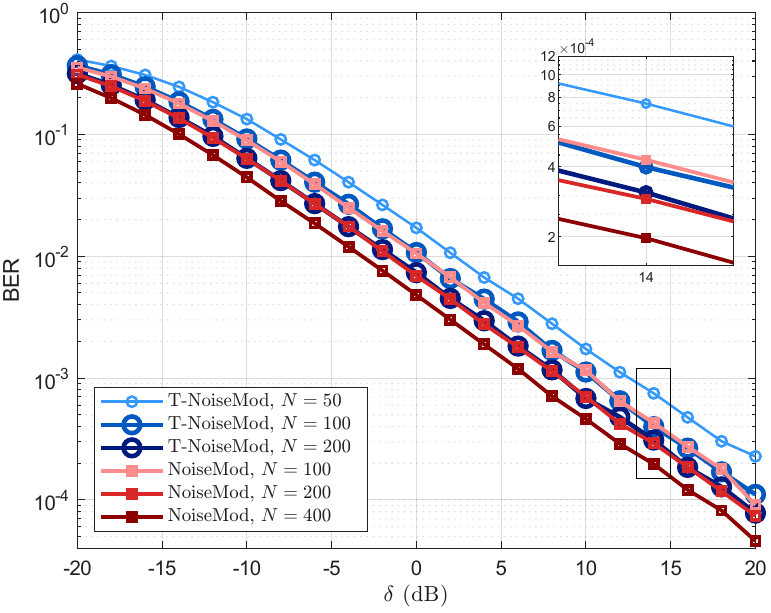}
    \vspace{-0.4cm}
    \caption{BER comparison of T-NoiseMod and conventional NoiseMod for varying observation lengths with $\alpha=10$, $\mu=1$.}
\vspace{-0.4cm}
    \label{fig:n_comp}
\end{figure}

As shown in Fig.~\ref{fig:n_comp}, increasing $N$ concentrates the sample mean and variance estimates around their true values, thereby reducing the BEP. The results demonstrate that binary NoiseMod achieves a slightly lower BEP for a given $N$, consistent with the fact that binary signaling involves fewer signaling levels than ternary signaling. However, the proposed T-NoiseMod conveys three information bits per two-block signaling interval, whereas binary NoiseMod conveys two bits over the same two-block interval. Therefore, even at operating points where BEP values are comparable, T-NoiseMod provides a higher information rate per signaling interval, enabling improved throughput with the same total number of transmitted blocks. Since both schemes operate on the same per-sample basis and occupy the same bandwidth, this rate advantage can also be viewed in terms of spectral efficiency, while the associated reliability trade-off is illustrated in Fig.~\ref{fig:n_comp}.

\vspace{-0.2cm}
\section{Conclusion}
In this letter, we have proposed T-NoiseMod, a ternary noise-based signaling scheme that embeds information in the statistics of artificial noise waveforms using three states: low-variance, high-variance, and silent. By pairing two consecutive length-$N$ blocks, eight valid state combinations are formed, enabling the transmission of three information bits per signaling pair. In the quasi-static Rayleigh fading channel with perfect CSI at the receiver, we have designed a two-stage detector that first performs sample-mean-based silent discrimination and then applies a variance-based LRT for low/high classification. Based on the preliminary single-block state transition probabilities and a pairwise consistency refinement that eliminates the illegal $(S,S)$ detector output, we have derived tractable approximate analytical BEP expressions by averaging the normalized Hamming distance over all valid final state pairs and then averaging over Rayleigh fading. Computer simulation results closely match the approximate theoretical analysis and confirm the expected spectral efficiency–reliability trade-off. By conveying three bits over two blocks, the proposed T-NoiseMod scheme improves the information rate over binary NoiseMod at the cost of a modest increase in BEP. Future work includes optimizing mapping and thresholds, extending the framework to imperfect CSI and multiuser scenarios, studying rate--reliability tradeoffs under practical power and bandwidth constraints, and investigating possible covertness properties under explicit warden models.
\vspace{-0.4cm}
\bibliographystyle{IEEEtran}
\bibliography{IEEEabrv,bib_2023}
\end{document}